\documentstyle[11pt,aaspp4]{article}

\received{4 August 1988}
\accepted{23 September 1988}

\begin{document}
\newcommand{\decsec}[2]{$#1\mbox{$''\mskip-7.6mu.\,$}#2$}
\newcommand{\decsectim}[2]{$#1\mbox{$^{\rm s}\mskip-6.3mu.\,$}#2$}

\title{CC Bo\"otis: QSO, Not Variable Halo Giant}

\author{Bruce Margon and Eric W. Deutsch}
\affil{Astronomy Department, University of Washington, Box 351580,
Seattle, WA 98195-1580 \\ Electronic mail: margon@astro.washington.edu, 
deutsch@astro.washington.edu}

\vskip .5in
\begin{center}
Accepted for publication in Pub. Astr. Soc. Pacific, Vol 109, June 1997
\end{center}
\vskip .3in

\begin{abstract}
The poorly-studied, faint ($18<m_{pg}<19.5$) variable star CC~Bo\"otis has
been noted in the literature as a candidate for a halo red giant. It proves
instead to be a quasi-stellar object of redshift $z=0.172$, and is detected
as an X-ray source by {\it ROSAT}. In addition to its odd heritage, CC~Boo
exhibits unusually high amplitude optical variability for an
optically-selected QSO.
\end{abstract}

\section{Introduction}

A faint variable star S~10762 was discovered on Tautenburg Schmidt plates in 
the M3 field near the North Galactic Pole by Meinunger (1972). On that plate 
material the object is seen to vary slowly and with no obvious periodicity in
the $18<m_{pg}<19.5$ range during the interval 1964--1972, and is noted as
``not red."  To our knowledge, no further observations of this object are
reported in the literature. Kukarkin et al. (1975) assign the object the name
CC~Bo\"otis, and it is cataloged in the 4th Edition of the General Catalog of
Variable Stars (\cite{kho85}) as a ``slow irregular variable." The next (and
only) mention of CC~Boo appears to be that of Jura and Kleinmann (1992), who
included the object in a list of candidate halo red giants, based on the GCVS
classification.

We became interested in CC~Boo when Jura (1997) noted its proximity to the 
{\it ROSAT} All Sky Survey Bright Source Catalog X-ray source
1RXS~J134021.4+274100 (Voges et al. 1997). As he points out, if the
identification with the X-ray source is correct, it
seems unlikely that the classification of the star as a faint red giant can be 
valid.

\clearpage
\section{Observations and Discussion}

A finding chart for CC~Boo is provided by Meinunger (1972), but for
convenience we provide in Figure~1 material more suited for use at a
large telescope, an image of the field obtained in the Gunn $r$-band
with the Astrophysical Research Consortium (ARC) 
3.5m telescope at Apache Point, New Mexico, using the Double Imaging
Spectrograph (DIS).  A dichroic mirror also provides $g$ images
simultaneously. We have measured the position of the star 
on an extraction from the Digitized Sky Survey, and derive
$\alpha_{2000}=13^{\rm h}40^{\rm m} \decsectim{22}{84},\
\delta_{2000}=27^\circ 40'$ \decsec{58}{6}. The formal internal errors of
our astrometric measurements
are negligible, but the external uncertainty of the entire reference frame is
normally of order $\pm1''$. Our coordinates differ from those previously
reported in Meinunger (1972) and the GCVS (in both cases truncated to the
nearest $1'$) by more than
$1'$, but this is probably not an alarming discrepancy; the
identification of the object is not in doubt due to the finding chart
in the discovery work. More to the point, the coordinates of the X-ray
source (Voges et al. 1997) are $\alpha_{2000}=13^{\rm h}40^{\rm m}
\decsectim{21}{4},\ \delta_{2000}=27^\circ 41' 01''$ ($1\sigma$ uncertainty
$11''$), an offset of $19''$ from our measured optical position, and thus
almost surely indicating that CC~Boo is the optical counterpart.

Via observations of standard stars from Landolt (1992) and photometric 
transformations 
from J\o rgensen (1994), we provide in Table 1 $V$ and $R$
data from our images for CC~Boo and several nearby objects which bracket it in
brightness. These may be useful for future monitoring of this highly variable 
object. Although the internal errors of the photometry are all $<0.02$~mag, 
these data should not be assumed more accurate than $\pm0.1$~mag due to 
systematic uncertainties in the color transformation.

We obtained the spectrum of CC~Boo with DIS on the 3.5m at low resolution
($\lambda/\Delta\lambda\sim400$) on UT 1997 January 22. Two exposures of
duration 600~s were obtained and co-added. An approximate flux calibration
was derived via observation of several spectrophotometric standards from
Massey et al. (1988) and Massey and Gronwall (1990); light losses at the slit
are such that these absolute fluxes cannot be regarded as more accurate than
$\pm0.3$~mag. However as the object is
known to be highly variable, accurate photometry is probably not an important
issue. The resulting spectrum is shown in Figure~2.

It is clear that CC~Boo is a low redshift quasi-stellar object or related AGN.
In addition to strong, broad H$\alpha$, H$\beta$, 
and H$\gamma$ emission, forbidden lines of
[O\,III] $\lambda\lambda$4959, 5007, [O\,II] $\lambda$3727, [Ne\,V] 
$\lambda$3426, and probably [Ne\,III] $\lambda$3869 are detected on a markedly
UV-excess continuum. From four narrow, well-detected forbidden lines, we 
derive $z=0.172\pm0.001$.

Optically-selected QSOs with variability amplitude as high as that of CC~Boo,
1.5~mag, are very rare. Thus this is an interesting object quite aside from
its obscure heritage. The issue of whether QSO variability amplitude is
correlated with luminosity, redshift, and selection technique is complex, and
there is still not total agreement in the literature on these points. However
by the standards of any of the studies, CC~Boo stands out. For example, Hook
et al. (1994) find {\it rms} variability amplitude of 0.25~mag for a large
sample of optically-selected QSOs with $M_B=-23$, similar to the object
discussed here.

Active galactic nuclei originally misclassified as variable stars are rare but
certainly not unprecedented. The prototype of such objects of course is 
BL~Lac. A close analog to CC~Boo is probably X~Com (Bond 1973, Bond and 
Sargent 1973), which has comparably high amplitude photometric variability 
(Green, Huchra, and Bond  1977).

We have searched for possible angular extension of the image of CC~Boo in our 
data, although the pixel scale of the DIS cameras is modest, 1.1 and 0.6
arcsec~pixel$^{-1}$ for the $g$ and $r$ images, respectively, and the seeing 
was poor during our observations. There is 
evidence for detection of an extended component in the $r$ 
image
CC~Boo: six nearby object of comparable magnitude have a mean image FWHM 
of $3.2\pm0.1$ pixels, while the QSO measures 4.0 pixels FWHM. A further 
attempt to resolve an underlying galaxy seems clearly worthwhile.

The X-ray count rate quoted by Voges et al. (1997) corresponds to a flux of 
$\sim(9\pm1.5)\times10^{-13}$~erg~cm$^{-2}$~s$^{-1}$, with uncertainties of up
to $\sim2\times$ possible due to the unknown X-ray spectral slope. For our
observed $V=17.8$, we infer {\it log} [$f_x/f_v$] = 0.44, very close to the mean
for AGN reported by Stocke et al. (1991). Thus CC~Boo has quite normal X-ray 
properties. The object lies in the survey region of, but is not detected in,
the MG~II (5~GHz) radio catalog (Langston et al. 1990), which is stated as
99\% complete at 70~mJy.

For the sake of completeness, we note the presence of the two faint
($m\sim21$), blue unconfirmed QSO candidates 1338.3+2756 and 1338.1+2759
(Crampton et al. 1985, 1988) located $3.5'$ and $3.4'$, respectively, from
the nominal {\it ROSAT} X-ray position. These objects are far beyond the quoted
X-ray positional uncertainties, and would also have extreme values of
[$f_x/f_v$] if identified with the X-ray source; we therefore believe them to
be unrelated to this discussion.
Those authors were unlucky not to have also discovered the nature of CC~Boo on 
their grens plates of this field; the red limit of their bandpass falls just 
shortward of the strong broad H$\beta$ emission.

In summary, CC~Boo appears to be a normal QSO in most
respects other than its originally cataloged classification. However, it 
is unusually highly variable for an optically-selected object, and for 
that reason alone probably warrants future monitoring, as well as radio 
observations.

\acknowledgments

We are very grateful to Dr. M. Jura for drawing our attention to this object.
The Digitized Sky Survey is
based on photographic data obtained using the Oschin Schmidt
Telescope on Palomar Mountain.  The Palomar Observatory Sky Survey was
funded by the National Geographic Society.  The Oschin  Schmidt
Telescope is operated by the California Institute of Technology
and Palomar Observatory.  The plates were processed into the present
compressed digital form with their permission.  The Digitized  Sky
Survey was produced at the Space Telescope Science Institute (STScI)
under U. S. Government grant NAG W-2166.

\clearpage

\begin{figure}
\plotone{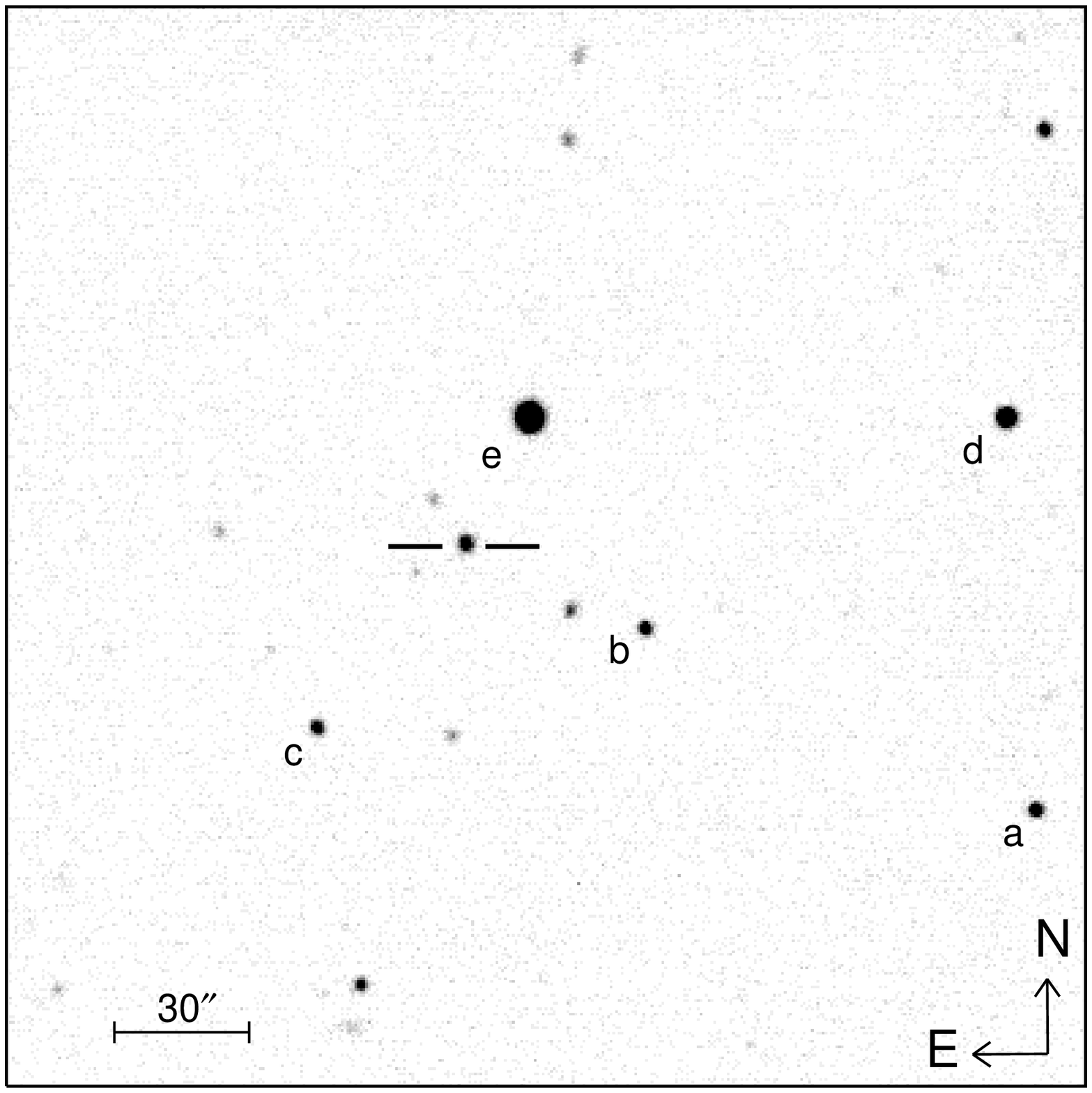}
\caption{A $4'\times4'$ Gunn $r$-band image of the field of CC~Boo. These data are a 
total of 90~s of integration summed from 3 exposures
obtained on UT 1997 January 25 with the ARC 3.5m telescope. The object is 
flagged, as are several nearby comparison stars selected to bracket CC~Boo in 
brightness. Photometry for these objects is given in Table 1. \label{fig1}}
\end{figure}

\begin{figure}
\plotone{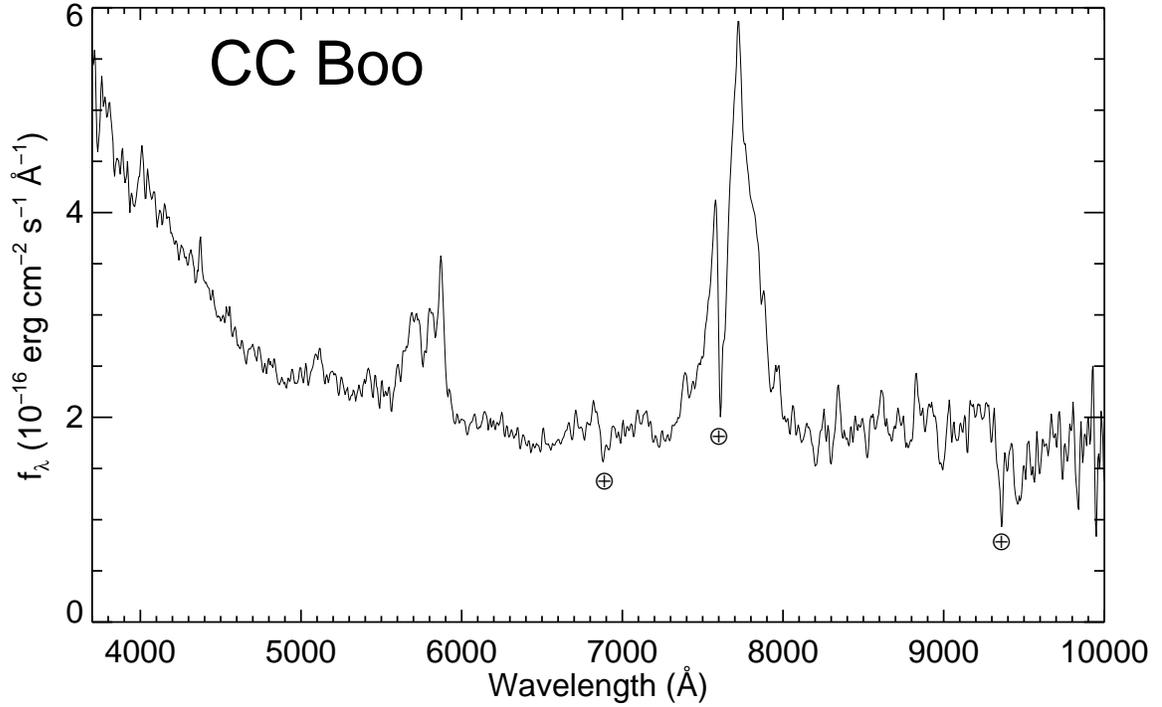}
\caption{The spectrum of CC~Boo. Emission line identifications are noted in
the text. The strong, broad H$\alpha$ emission is mutilated by the telluric
A-band. \label{fig2}}
\end{figure}

\begin{deluxetable}{ccc}
\tablenum{1}
\tablecolumns{3}
\tablewidth{370pt}
\tablecaption{Photometry in the CC Boo field\tablenotemark{a}}
\tablehead{
\colhead{Star\tablenotemark{b}} &
\colhead{$V$} &
\colhead{$V-R$}
}
\startdata
CC Boo & 17.79 & 0.34 \nl
a & 17.78 & 0.40 \nl
b & 18.03 & 0.43 \nl
c & 18.29 & 0.51 \nl
d & 16.24 & 0.34 \nl
e & 14.53 & 0.40 \nl
\enddata
\oddsidemargin -0.7in
\textwidth 7.3in
\tablenotetext{a}{\,Internal accuracy of the photometry is $<0.02$~mag, but 
systematic uncertainties of $\pm$0.1~mag are likely due to color 
transformations.}
\tablenotetext{b}{\,Designations $a$, $b$, and $c$ identical to the nomenclature
of Meinunger (1972).}
\end{deluxetable}

\end{document}